# Advances in Macromolecular Data Storage

Masud Mansuripur

College of Optical Sciences, The University of Arizona, Tucson, Arizona 85721

**Abstract**. We propose to develop a new method of information storage to replace magnetic hard disk drives and other instruments of secondary/backup data storage. The proposed method stores petabytes of user-data in a sugar cube (1 cm$^3$), and can read/write that information at hundreds of megabits/sec. Digital information is recorded and stored in the form of a long macromolecule consisting of at least two bases, $A$ and $B$. (This would be similar to DNA strands constructed from the four nucleic acids $G, C, A, T$.) The macromolecules initially enter the system as blank slates. A macromolecule with, say, 10,000 identical bases in the form of $AAAAA....AAA$ may be used to record a kilobyte block of user-data (including modulation and error-correction coding), although, in this blank state, it can only represent the null sequence 00000....000. Suppose this blank string of $A$'s is dragged before an atomically-sharp needle of a scanning tunneling microscope (STM). When electric pulses are applied to the needle in accordance with the sequence of 0s and 1s of a 1 $kB$ block of user-data, selected $A$ molecules will be transformed into $B$ molecules (e.g., a fraction of $A$ will be broken off and discarded). The resulting string now encodes the user-data in the form of $AABABBA...BAB$. The same STM needle can subsequently read the recorded information, as $A$ and $B$ would produce different electric signals when the strand passes under the needle. The macromolecule now represents a data block to be stored in a "parking lot" within the sugar cube, and later brought to a read station on demand. Millions of parking spots and thousands of Read/Write stations may be integrated within the micro-fabricated sugar cube, thus providing access to petabytes of user-data in a scheme that benefits from the massive parallelism of thousands of Read/Write stations within the same three-dimensionally micro-structured device.

**1. Introduction**. Many of the traditional problems in disk and tape data storage can be overcome if data-blocks were released from the confines of a disk (or tape) and allowed to float freely between Read/Write stations (i.e., heads) and permanent parking spots; see Fig. 1. The heads and parking spots thus become fixed structures within an integrated chip, while the macromolecular data blocks themselves become the (mobile) storage media. In this scheme, a large number of Read/Write heads could operate in parallel, the heads and parking spots would be constructed (layer upon layer) in a truly three-dimensional fashion, and individual nanometer-sized molecules—strung together in a flexible macromolecular chain—would be used to represent the 0s and 1s of binary information. We have discussed the potential advantages of this alternative scheme for secondary data storage in previous publications[1,2] and, to demonstrate feasibility, conducted preliminary experiments using short DNA strands that reside and travel within micro-fluidic chambers.[2-6] Since our first proposal in 2001 to develop macromolecular data storage systems,[1] other groups have contributed to developments in DNA-based techniques, with ramifications for information storage/processing. In this paper we review the basic principles of macromolecular data storage, and also briefly describe the latest developments that could impact the implementation of a practical and commercially viable storage system.

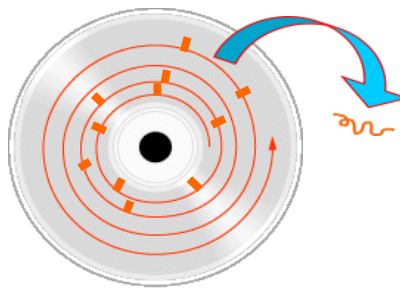

**Fig. 1**. Releasing data-blocks from the confines of the storage medium and giving them an independent physical identity solves many of the traditional problems associated with disk/tape data storage. One will no longer be limited to surface (i.e., 2D) data storage. Also, Read/Write heads can become stationary, thus enabling the integration of numerous heads into a single chip.

**2. Device Architecture**. As a possible alternative to present-day mass data storage devices (e.g., magnetic and optical disks and tapes, flash memories), we envision a system in which data blocks are encoded into macromolecules constructed from two (or more) distinct bases, say, $A$ and $B$; the bases can be strung together in arbitrary order such as $AABABBAB \ldots ABA$ to represent binary sequences of user-data ($A = 0, B = 1$).[1,2] The macromolecular data blocks must be created in a *Write station*, transferred to *parking spots* for temporary storage, and brought to a *Read station* for decoding and readout. The *erase* operation is as simple as discarding a data block and allocating its parking spot to another macromolecule. (In principle, discarded molecules can be recycled after being broken down to their constituent elements.) The parking spots and Read/Write stations depicted in Fig. 2, for example, are microfluidic chambers connected via micro-channels and micro-valves, as shown in Fig. 3, which enable automatic access through an electronic addressing scheme.[2] With the dimensions of the various chambers indicated in Fig. 2, one can readily incorporate, on a $1.0\ cm^2$ surface area, a total of $10^6$ parking spots ($\sim 0.25\ cm^2$), 1000 read/write stations ($\sim 0.1\ cm^2$), and necessary plumbing (e.g., $1\ \mu m$-wide connecting routes, $1 \times 1\ \mu m^2$ binary valves or switches), which would occupy an area $\sim 0.65\ cm^2$. Assuming megabyte-long data-blocks, the storage capacity of the $10^6$ parking spots in this scheme will be $10^{12}\ bytes/cm^2$. In a 3D design based on $10\ \mu m$-thick layers, the capacity of the proposed macromolecular storage system would exceed $10^{15}\ bytes/cm^3$. For a comparison with a current state-of-the-art technology, note that storing $10^{15}$ bytes of data on Digital Versatile Disks requires a 128 meter-tall stack of $12\ cm$-diameter DVD platters.

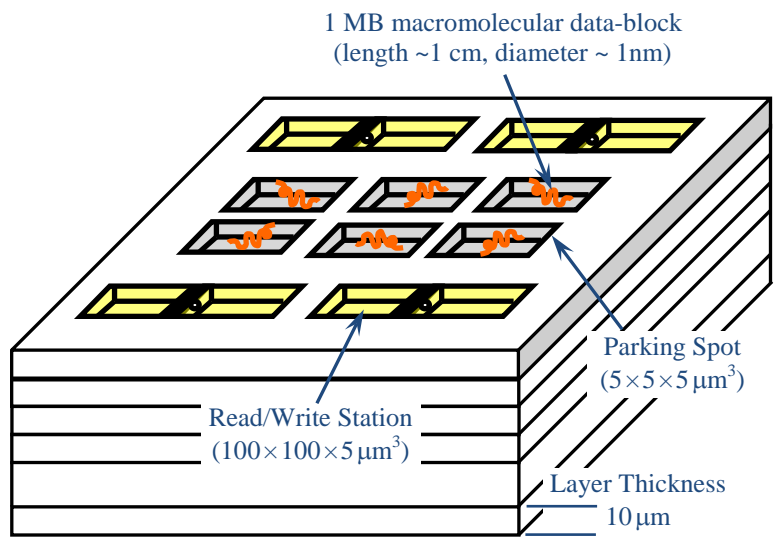

**Fig. 2**. Chip surface utilization. The same arrangement of Read/Write stations and parking spots is repeated in stacked layers.

The schematic in Fig. 3 shows the patterned surface of one possible embodiment of the proposed data storage chip. Once a parking spot is selected, its macromolecular content will be transferred to the Read/Write station under the influence of an applied electric voltage (i.e., electrophoretic transfer). Following the completion of a read/write operation, the molecular strand returns to its designated parking spot. The respectable data-rates of individual Read/Write stations (conservatively estimated to be on the order of a few megabits/sec) are augmented by the massive parallelism of thousands of Read/Write stations integrated on the same micro-chip.

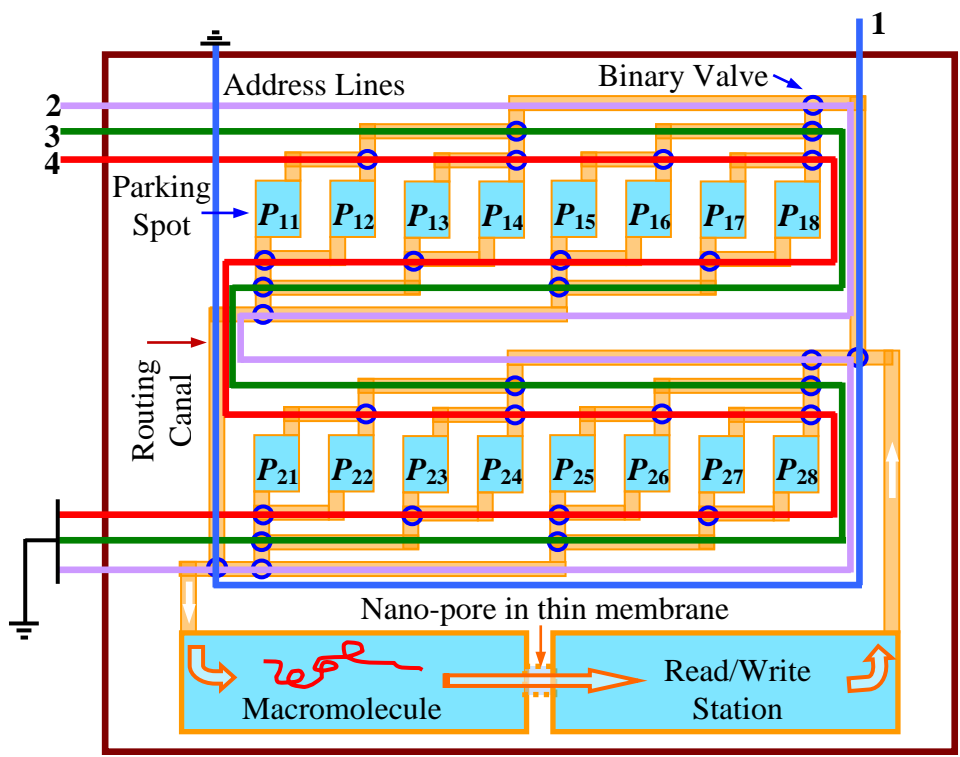

**Fig. 3**. Example of a microfluidic system containing 16 parking spots, one Read/Write station, and transport channels (with electronically controlled micro-valves) for macromolecular data storage.

One may also envision packaging the molecular data blocks in microscopic spherical vesicles known as liposomes, as shown in Fig. 4. This would provide a robust environment for the transfer of data blocks between the parking spots and the Read/Write stations. Moreover, by planting specific antibody molecules on the vesicle surfaces, it becomes possible to direct these *addressed packages* to desired locations tagged with antibody-specific antigens (i.e., receptors).

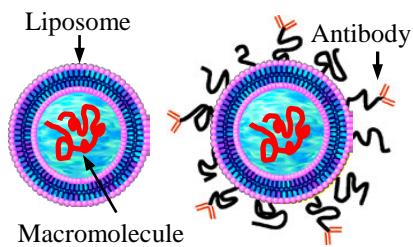

**Fig. 4**. Liposomes are microscopic spherical vesicles that form when phospholipids are hydrated. When mixed in water under low shear conditions, the phospholipids arrange themselves in sheets, the molecules aligning side by side with heads up and tails down. The sheets then join tails-to-tails to form a bilayer membrane. Phospholipid bilayer spheres enclosing water and one or more macromolecules can be produced with uniform size (diameter ~ 200 nm or smaller). Antibodies inserted in the outer surface of the liposome can target specific locations within the device tagged with antibody-specific antigens.

**3. Recent Advances in Nanopore-based Macromolecular Processing**. Macromolecules can be readily fabricated as fixed-length homopolymers of arbitrary length. If each base of the homopolymer is denoted by *A*, the initial form of each such macromolecule will be $AAAA\ldots AA$, representing the $0000\ldots00$ (null) binary sequence. An atomically-sharp needle, such as the probe used in a scanning tunneling microscope (STM) may subsequently be brought into contact with (or in close proximity to) the null macromolecule in order to modify some of its bases—say, by breaking off a section of a given base *A* and turning it into a smaller base *B*. By selectively applying an electric pulse to the STM needle, one may thus create a tailored macromolecule *AABAB…BBA* representing the desired binary sequence.

Dragging a macromolecule in front of the STM needle can be done in a controlled fashion using, for example, the *DNA Transistor* described by IBM researchers in recent years.[7] This "transistor," which is shown schematically in Fig.5(a), consists of two micro-chambers connected by a nano-pore. The metallic electrodes embedded in the nano-pore walls are used to deliver highly-localized electrical signals to the translocating DNA strand. Figure 5(b) is a magnified view of the macromolecule (yellow strand) moving through a solid-state nano-pore articulated with probes. These probes provide a controlled and localized electric field and/or electrical current (just as an STM needle would) to modify the local properties of the molecule.

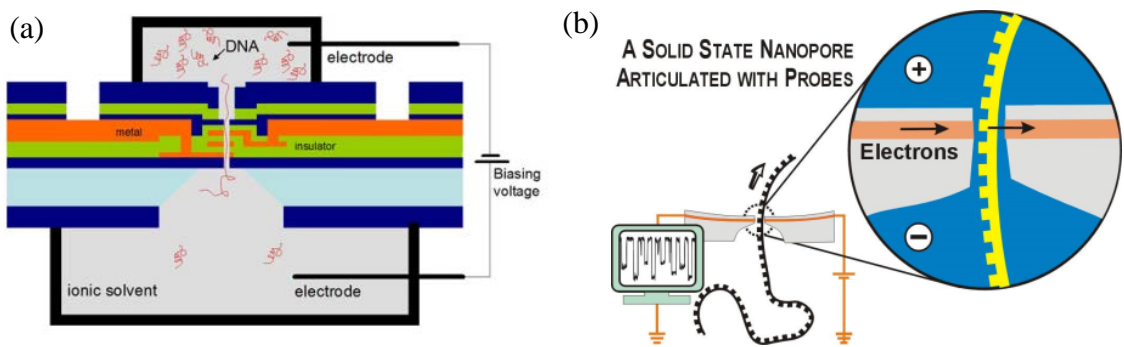

**Fig.5**. (a) IBM's DNA "transistor." The biasing voltage induces strands of DNA to pass through the nano-pore. Ordinarily, DNA would move through the nano-pore at a speed of several million bases per second were it not for metal contacts inside the nano-pore that create an internal electric field to trap the DNA. By alternating the voltages applied to the metal contacts, DNA is ratcheted from the top reservoir to the bottom reservoir in a controlled manner.[7] (b) Close-up view, showing the electronic reading/writing of a macromolecular strand using articulated nano-pore probes.

Decoding DNA strands by translocation through biological nano-pores is a fairly old idea, which has been investigated by many researchers for at least the past two decades.[8-16] More recently, progress has been made in fabricating solid-state nano-pores in free-standing thin dielectric films or graphene membranes in conjunction with focused electron- or ion-beam micro-machining;[17] see Fig.6. We believe the tools and techniques developed in recent years for DNA sequencing can be adapted to carry out not only readout at high data-rates, but also high-speed recording of information onto macromolecular strands representing blocks of binary user-data in our proposed scheme of macromolecular data storage.

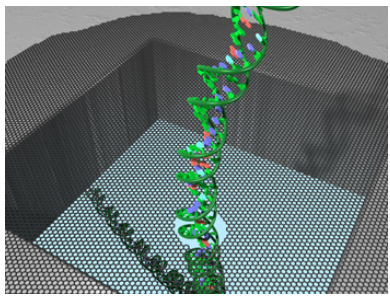

**Fig.6**. Schematic of a graphene membrane hosting a nano-pore drilled with the aid of a sharply-focused electron beam.[17] Also shown is a double-stranded DNA molecule translocating through the nano-pore.

**4. DNA Data Storage Finally Attracts Attention**. Our thirteen-year-old ideas have been rediscovered in the past couple of years! The August 20, 2012 issue of *Time* magazine had an article titled "*The First Book To Be Encoded in DNA: Two Harvard scientists have produced 70 billion copies of a book in DNA code—and it's smaller than the size of your thumbnail.*" The reference is to a paper that was to be published in *Science* later that year.[18] Then, on January 26, 2013, *The Economist* published an article with the headline: "*Archives could last for thousands of years when stored in DNA instead of magnetic tapes and hard drives.*" According to this article, two scientists have discovered a "new scheme" and "managed to set a record (739.3 kilobytes) for the amount of unique information encoded" in synthetic DNA strands. This time the focus of attention was the work of two other scientists at the European Bioinformatics Institute, which was to be reported in an upcoming issue of *Nature*.[19] (See also a related article in *The New York Times* of January 28, 2013,[20] and the lead editorial in the *American Journal of Neuroradiology*, January 2014.[21]) The claim is that this "new scheme" should be "capable of swallowing the roughly 3 zettabytes ($10^{21}$ bytes) of digital data thought presently to exist in the world and still have room for plenty more." The estimated storage density of the "new scheme," according to *The Economist*, is "around 2.2 petabytes ($10^{15}$) per gram." Of course, a quick glance through our papers[1-6] dating back to 2001 through 2007 reveals that we had foreseen this potential of DNA for high-density data storage a decade earlier. The use of modulation coding and error-correction coding, which is now heralded as a novel feature of the "new scheme," was also implicit in our early papers because our macromolecular strands were equated with blocks of user-data recorded as individual sectors on optical disks and magnetic hard disks, where the use of modulation and error-correction coding is essential for the fidelity of the stored data.

Whereas our papers[2-6] described techniques and structures for rapid reading/writing of information in macromolecular media, in general, and DNA strands, in particular, the schemes proposed by others in the past couple of years do *not* go beyond the standard methods of synthesizing and sequencing DNA molecules. Here again is *The Economist*:

> "The researchers were able to encode and decode five computer files, including an MP3 recording of part of Martin Luther King's 'I have a dream' speech and a PDF version of the 1953 paper by Francis Crick and James Watson describing the structure of DNA."

> "The simplest approach would be to synthesize one long DNA string for every file to be stored. But DNA-synthesis machines are not yet able to do that reliably. So the researchers decided to chop their files into thousands of individual chunks, each 117 bases long. In each chunk, 100 bases are devoted to the file data themselves, and the remainder used for indexing information that records where in the completed file a specific chunk belongs. The process also contains the DNA equivalent of the error-detecting 'parity bit' found in most computer systems."

> "Reading the chunks back is simply a matter of generating multiple copies of the fragments using a standard chemical reaction, feeding these into a DNA-sequencing machine and stitching the files back together."

> "There are downsides to DNA as a data-storage medium. One is the relatively slow speed at which data can be read back. It took the researchers two weeks to reconstruct their five files, although with better equipment it could be done in a day."

Needless to say, our proposed schemes of macromolecular data storage,[1-6] which predate the "new scheme" by nearly a decade, are fast, automated, reliable, and capable of reading, writing, and erasing at speeds that are comparable to those of optical and magnetic disk drives.

**5. Concluding Remarks**. We have proposed macromolecular media and related techniques for 3D storage of massive amounts of binary information. The proposed system architecture is in the form of stacked layers containing numerous parking lots and Read/Write stations for the storage and processing of molecular strands into which user-data is encoded. Storage densities start at petabytes/cm³ (i.e., $10^{15}$ bytes of user-data in a sugar cube), and have the potential to grow by another five orders of magnitude to $10^{20}\ bytes/cm^3$. Although individual Read/Write channels are expected to be relatively slow (having speeds probably on the order of a few $megabits/sec$), the massive parallelism inherent in our proposed scheme would allow the operation of thousands of Read/Write stations in parallel, thereby achieving data-transfer rates that would be superior to those available in conventional storage systems. Existing bio-chemical techniques can be adapted and extended to perform read/write operations within the integrated environment of a microchip. Device fabrication entails micro- and nano-fabrication, micro-fluidics, optical and/or electronic manipulation of small objects (e.g., optical tweezers), and bio-chemical nano-processing.